\documentclass[conference]{IEEEtran}
\IEEEoverridecommandlockouts
\usepackage{cite}
\usepackage{amsmath,amssymb,amsfonts}
\usepackage{algorithmic}
\usepackage{graphicx}
\usepackage{textcomp}
\usepackage{xcolor}
\usepackage{amsmath,graphicx, siunitx, graphicx,subcaption,multicol, tabularx, multirow, array}

\usepackage{adjustbox}
\usepackage{pgfplots}
\usepackage{diagbox}
\usepackage{caption}
\usepackage{pgfplots}
\usepackage{diagbox}
\usepackage{caption}
\usepackage{booktabs}
\usepackage{filecontents}

\def\BibTeX{{\rm B\kern-.05em{\sc i\kern-.025em b}\kern-.08em
    T\kern-.1667em\lower.7ex\hbox{E}\kern-.125emX}}
\begin{document}

\title{Audio Data Augmentation for Acoustic-to-articulatory Speech Inversion
\thanks{This work was supported by the National Science Foundation grant IIS1764010.}}

\author{\IEEEauthorblockN{1\textsuperscript{st} Yashish M. Siriwardena}
\IEEEauthorblockA{\textit{Department of Electrical and Computer Engineering} \\
\textit{University of Maryland College Park, MD, USA}\\
yashish@umd.edu}
\and
\IEEEauthorblockN{2\textsuperscript{nd} Ahmed Adel Attia}
\IEEEauthorblockA{\textit{Department of Electrical and Computer Engineering} \\
\textit{University of Maryland College Park, MD, USA}\\
aadel@umd.edu}
\and
\IEEEauthorblockN{3\textsuperscript{rd} Ganesh Sivaraman}
\IEEEauthorblockA{
\textit{Pindrop, GA, USA}\\
ganesa90@gmail.com}
\and
\IEEEauthorblockN{4\textsuperscript{th} Carol Espy-Wilson}
\IEEEauthorblockA{\textit{Department of Electrical and Computer Engineering} \\
\textit{University of Maryland College Park, MD, USA}\\
espy@umd.edu }
}

\maketitle

\begin{abstract}
Data augmentation has proven to be a promising prospect in improving the performance of deep learning models by adding variability to training data. In previous work with developing a noise robust acoustic-to-articulatory speech inversion (SI) system, we have shown the importance of noise augmentation to improve the performance of speech inversion in 'noisy' speech conditions. In this work, we extend this idea of data augmentation to improve the SI systems on both the clean speech and noisy speech data by experimenting three data augmentation methods. We also propose a Bidirectional Gated Recurrent Neural Network as the speech inversion system instead of the previously used feed forward neural network. The inversion system uses mel-frequency cepstral coefficients (MFCCs) as the input acoustic features and six vocal tract-variables (TVs) as the output articulatory targets. The Performance of the system was measured by computing the correlation between estimated and actual TVs on the Wisconsin X-ray microbeam database. The proposed speech inversion system shows a 5\% relative improvement in correlation over the baseline noise robust system for clean speech data. The pre-trained model, when adapted to each unseen speaker in the test set, improves the average correlation by another 6\%. 

\end{abstract}


\section{Introduction}

Acoustic-to-articulatory speech inversion (SI) is the task of retrieving the articulatory dynamics for a given speech signal \cite{Sivaraman_ASA}. The mapping from acoustics to articulation is an ill-posed problem which is known to be highly non-linear and non-unique \cite{Qin2007}. However, over the recent years, the development of Speech Inversion (SI) systems has gained attention due to its potential in wide range of applications ranging from Automatic Speech Recognition (ASR) \cite{Mitra2011_new}, speech synthesis \cite{speech_synthesis_1}, speech therapy \cite{Fagel2008A3V} and mental health assessments \cite{espywilson19_interspeech, Siriwardena_SZ}. \cite{espywilson19_interspeech,Siriwardena_SZ}. Real articulatory data is obtained using techniques like X-ray microbeam \cite{Westbury1994a}, Electromagnetic Articulometry (EMA) \cite{Tiede2017} and real-time Magnetic Resonance Imaging (rt-MRI) \cite{Narayanan2004}. These techniques are expensive, time consuming and require specialized equipment for observing articulatory movements directly \cite{Sivaraman_ASA}. Hence, developing a speaker-independent SI system which can accurately estimate articulatory features for any unseen speaker can transform how speech research is conducted. 

Advancements in deep neural networks (DNNs), especially in processing time series data to capture contextual information has propelled the development of SI systems to new heights. Bidirectional LSTMs (BiLSTMS) \cite{illa18_interspeech}, CNN-BiLSTMs \cite{Shahrebabaki2020}, Temporal Convolutional Networks (TCN) \cite{shahrebabaki21_interspeech} and transformer models \cite{udupa21_interspeech} have gained state-of-the-art results with multiple articulatory datasets \cite{Tiede2017}. In our previous work, we have reported SI results with the XRMB dataset \cite{Sivaraman_ASA} where a simple feed-forward neural network was trained with manually contextualized MFCCs as input features. In this work, we propose a Bidirectional Gated Recurrent Neural Network (BiGRNN) which outperforms the existing feed-forward neural network model on predicting TVs with the XRMB dataset. The BiGRNN model is light-weight, does not need any pre-contextualization of the input audio features or post-processing with low pass filters.

Majority of the SI systems are usually trained with a single corpus of data and have shown to perform poorly in cross-corpus \cite{seneviratne19_multicorpus} or speaker-independent experiments \cite{Shahrebabaki2020}. Difficulty collecting a larger corpus of data with many subjects and the differences in procedures of articulatory data collection (placement of sensors, EMA vs XRMB) have also made things complicated. To address the issue of data scarcity, the machine learning community has recently turned in the direction of data augmentation and generating synthetic data to train DNN models. The idea of training a DNN model on similar but different examples is known as data augmentation, which was initially proposed in \cite{Simard2012} and later formalized by the Vicinal Risk Minimization (VRM) principle in \cite{Chapelle_VRM}. Data augmentation has been widely utilized in improving the robustness of speech applications over the last few years (eg. automatic speech recognition (ASR)\cite{ASR_dataaug_2} and speech emotion recognition \cite{SER_dataaugment}). However, limited work has been done in the articlulatory speech inversion domain, specially to improve the overall performance of SI systems with different, readily available data augmentation techniques. In previous work with developing a noise robust articulatory SI system, we explored the idea of generating additive noise audio data \cite{seneviratne18_interspeech}. The goal of that work was to develop a SI system which is robust against noisy acoustic data which is a common issue with rt-MRI articulatory datasets that contain significant amounts of MRI machine noise. Compared to the SI system trained entirely with clean speech data, the noise robust model did not perform well with the clean speech test set. The idea of developing a SI system which is robust against noisy speech data while also outperforming the current best performing model on clean speech data is the impetus for the current study.


\vspace*{-5pt}
\section{Dataset Description}
\vspace*{-3pt}

\subsection{X-Ray Microbeam (XRMB) dataset}
\vspace*{-2pt}

The original University of Wisconsin XRMB database \cite{Westbury1994a} comprises of naturally spoken isolated sentences and short read paragraphs collected from 32 male and 25 female subjects. These speech utterances were recorded along with trajectory data captured by X-ray microbeam cinematography of the midsagittal plane of the vocal tract using pellets placed on several articulators: upper (UL) and lower (LL) lip, tongue tip (T1), tongue blade (T2), tongue dorsum (T3), tongue root (T4), mandible incisor (MANi), and (parasagittally placed) mandible molar (MANm). However some of the articulatory recordings were marked as mistracked in the database and eliminating these samples left us with 46 speakers (21 males and 25 females) with a total of around 4 hours of speech data.

The anatomy of the speaker's vocal tract defines the absolute positions of the articulators. Since the X-Y positions of the pellets strongly depend on the anatomy of the speakers and variability of pellet placements, the measurements can vary significantly across speakers. Hence, to better represent vocal tract shape, relative measures were used to calculate the Tract Variables (TVs) from the X-Y positions of the pellets. TVs lead to a relatively speaker independent representation of speech articulation and characterize salient features of the vocal tract area function \cite{McGowan1994}. The TVs are based on articulatory phonology, a theoretical framework for speech production \cite{Browman1992}. Using geometric transformations, the XRMB trajectories were converted to TV trajectories as outlined in \cite{Mitra2012}. The transformed XRMB database comprises of six TV trajectories: Lip Aperture (LA), Lip Protrusion (LP), Tongue Body Constriction Location (TBCL), Tongue Body Constriction Degree (TBCD), Tongue Tip Constriction Location (TTCL) and, Tongue Tip Constriction Degree (TTCD).

\vspace*{-5pt}
\section{Audio Data Augmentation}
\label{data_augmentation}
\vspace*{-4pt}

We conducted 3 types of audio data augmentations using the Audiomentation library\footnote[1]{https://github.com/iver56/audiomentations}.

\vspace*{-6pt}
\subsection{Background Noise and Music}
\vspace*{-4pt}
\label{data_aug}

We added the noise and music data from MUSAN corpus \cite{musan2015} with the audio files in the XRMB dataset. The music files from the corpus were manually checked to remove any files with singing voices/human voices. We created two copies of the original speech data, one adding noise and the other adding music. The gain of noise and music added are set to be proportional to the Root Mean Square (RMS) value of the input sound in the audio. The added noise/music can range from 5dB to 20dB SNR and the level is randomly chosen.  


\vspace*{-10pt}
\subsection{Gaussian Noise}
\vspace*{-4pt}

Two copies of the original audio file are generated by adding Gaussian noise ranging from 5dB SNR to 20dB SNR. The noise SNR is sampled uniformly from the [5dB, 20dB] range when creating the noisy copy of the data.  

\vspace*{-10pt}
\subsection{Environmental Impulse Response (IR) functions}
\vspace*{-4pt}

To add reverberation noise, we used the environmental IR functions from the MIT Acoustical Reverberation Scene Statistics Survey corpus \cite{MIT_IR}. Similar to the previous noise types, two noise copies of the original audio files were generated by convolving with randomly chosen IR functions from the corpus. 

\vspace*{-10pt}
\section{Speech Inversion Systems}

\vspace*{-2pt}
\subsection{Input Audio Features}
\vspace*{-2pt}

All the audio files are first segmented to 2 second long segments and the shorter ones are zero padded at the end. Mel-Frequency Cepstral Coefficients (MFCCs) and Melspectrogram (MSPEC) features are then extracted as the acoustic input features for SI systems. Both MFCCs and MSPECs were extracted using a 20ms Hamming analysis window with a 10ms frame shift. For MFCCs, 13 cepstral coefficients were extracted for each frame while 40 Mel frequencies were used for both MFCCs and MSPECs. $\Delta$ and $\Delta \Delta$'s for MFCCs were also computed to be used as an extended input feature set along with MFCCs. Both MFCCs (and $\Delta$s) and MSPECs are utterance wise normalized (z-normalized) prior to model training. Table \ref{table: clean_train_clean_test} shows the results for how well each audio feature type performed with the SI systems.

\vspace*{-8pt}
\subsection{Proposed Bidirectional Gated RNN (BiGRNN) Model}

\begin{figure} 
    \centering
    \includegraphics[scale=0.37]{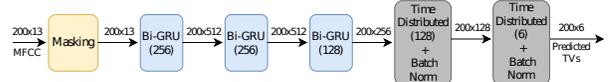}
    \caption{Proposed BiGRNN model architecture}
    \label{fig:model_archi}
\end{figure}

Gated Recurrent Unit (GRU) was first proposed by Cho et al. \cite{cho-etal-2014-properties} and has only 2 gates compared to 3 gates in a conventional LSTM unit, hence resulting in relatively smaller models and takes lesser time for training. In this paper, we propose a Bidirectional Gated RNN model as the speech inversion system. The model has 3 bidirectional layers of Gated Recurrent Units (GRUs) followed by two time distributed fully connected layers. Figure \ref{fig:model_archi} shows the detailed model architecture used with the data augmented XRMB dataset. We specifically used a Masking layer at the input to avoid the affect of padded zeros for TV predictions. Dropout layers were used after every layer to minimize over-fitting.

\vspace*{-8pt}
\subsection{Baseline BiLSTM and CNN-BiLSTM Models}
\vspace*{-2pt}

To compare against the proposed BiGRNN model, we used a BiLSTM model inspired by the work of \cite{illa18_interspeech} and a CNN-BiLSTM model similar to that in \cite{Shahrebabaki2020} as the baseline models. The BiLSTM model was trained with MFCCs and the CNN-BiLSTM model was trained with MSPECs. 

The BiLSTM model has 3 bidirectional LSTM layers followed by two time distributed fully connected layers. The model has the same architecture as the proposed BiGRNN model except that it uses BiLSTM layers instead of the BiGRNN layers in the front-end of the model. The CNN-BiLSTM model consists of 5 CNN layers, whose outputs are then concatenated together and fed to 2 BiLSTM layers, followed by a final CNN layer. 

All the models are evaluated with Mean Squared Error (MSE), Mean Absolute Error (MAE) and Pearson Product Moment Correlation (PPMC) scores computed between the estimated TVs and the corresponding ground-truth TVs. 


\vspace*{-5pt}
\subsection{Model Training}
\vspace*{-2pt}

The input XRMB dataset was divided into training, development, and testing sets, so that the training set has utterances from 36 speakers and the development and testing sets have 5 speakers each (3 males,2 females). None of the training, development and testing sets have overlapping speakers and hence all the models are trained in a `speaker-independent' fashion. The split also ensured that around 80\% of the total number of utterances were present in the training and the development and testing sets have a nearly equal number of utterances. This allocation was done in a completely random manner. When training the models with data augmentations as discussed in section \ref{data_augmentation}, all the augmented audio files were included in the same split as its original audio file to preserve the `speaker-independence' and to not affect the original train-dev-test splitting ratios. 


All the models were implemented with TensorFlow-Keras machine learning framework and trained with NVIDIA TITAN X GPUs. MSE, MAE and PPMC were experimented as loss functions to optimize the models. PPMC loss is defined as $1-PPMC$ and was used as a custom loss to optimize the models. Table \ref{table_loss_funcs} shows the average correlation on the test set for the best performing models with the 3 types of loss functions. It can be seen that the BiLSTM and BiGRNN models converged better with MAE loss where as CNN-BiLSTM model worked better with the PPMC loss. One limitation with the PPMC loss is that it could predict TVs with an offset from the ground-truth which is expected (since it only captures variability more than the actual values). To address that we combined PPMC loss with the MAE loss where a weight ($0<\alpha<1$) was assigned to PPMC loss and $1-\alpha$ was assigned to MAE loss. All the CNN-BiLSTM models are trained with the new weighted loss function which outperformed the MSE and MAE losses.\\

\vspace*{-10pt}
\begin{table}[th]
  \caption{Comparison of Loss functions with each Model type}
  \vspace*{-5pt}
  \centering
  \resizebox{\columnwidth}{!}{
  \begin{tabular}{|c |c |c |c |c |c|}
    \hline
    \multicolumn{1}{| c |}{\backslashbox{\textbf{Model}}{\textbf{Loss Fn.}}} & 
                        \textbf{MSE}  &
                        \textbf{MAE}  &
                        \textbf{PPMC}  & \multicolumn{1}{c|}{ \begin{tabular}{@{}c@{}}\textbf{Weighted Loss} \\ \textbf{($\alpha=0.8$)} \end{tabular}
                        }  &  \multicolumn{1}{c|}{ \begin{tabular}{@{}c@{}}\textbf{Weighted Loss} \\ \textbf{($\alpha=0.5$)} \end{tabular}
                        } \\
    \hline
    \label{table: augmentation_comparison}
    \textbf{BiGRNN} & 0.7910 & \textbf{0.7959} & 0.7917 & 0.7928 & 0.7940\\
    \textbf{BiLSTM} & 0.7850 & \textbf{0.7870} & 0.7858 & 0.7862 & 0.7865\\
    \textbf{CNN-BiLSTM} & 0.7163 & 0.7214 & \textbf{0.7333} & 0.7256 & 0.7203
    \label{table_loss_funcs}
    \\\hline

  \end{tabular}}
\end{table}
\vspace*{-2pt}

ADAM optimizer with a starting learning rate of 1e-3 and an exponential learning rate scheduler was used. The starting learning rate was maintained up to 20 epochs and then decayed exponentially after each subsequent 5 epochs. The model was trained with an early stopping criteria (patience=5) monitoring the `validation loss' on the development set. To choose the best starting `learning rate' (LR), we did a grid search on [1e-3, 3e-4, 1e-4], whereas to choose the training batch size we did a similar grid search on [16,32,64,128]. Based on the validation loss, 1e-3 and 128 were chosen as the learning rate and batch size respectively.

\vspace*{-3pt}
\section{Experiments and Results}
\vspace*{-3pt}

\subsection{Performance comparison of different model architectures with different audio features}
\vspace*{-2pt}

Previous work with developing SI systems have shown MFCCs to be superior over MSPECs and Perceptual Linear Predictions (PLPs) as acoustic features \cite{Sivaraman_ASA}. Our results with different audio features as shown in Table \ref{table: clean_train_clean_test} are consistent with the previous studies which suggests MFCCs can be effective specially with Bidirectional RNN based models. We also show that adding $\Delta$ and $\Delta\Delta$ features derived from MFCCs does not necessarily improve the performance. Table \ref{table: clean_train_clean_test} also presents the results of how different model architectures when trained with `clean speech' data performs on unseen speaker data. The BiGRNN model had a clear advantage over the two baseline models (BiLSTM and CNN-BiLSTM) and hence was selected as the proposed model architecture for the SI system. It should also be noted, that in this experiment, all the models were trained with only clean speech data (with no data augmentation) to disentangle any affects of data augmentation to choose the best DNN architecture.

\vspace*{-3pt}
\begin{table}[th]
  \caption{PPMC scores for models trained with clean speech when tested with clean speech test set}
  \vspace*{-3pt}
  \centering
  \label{table: clean_train_clean_test}
  \resizebox{\columnwidth}{!}
  {\begin{tabular}{|l|l|l|l|l|l|l|l|l|}
    \hline
    \multicolumn{1}{|l|}{\textbf{Model}} & 
                                    \multicolumn{1}{|l|}{\textbf{Audio features}}  &      \multicolumn{1}{|l|}{\textbf{LA}}  &  \multicolumn{1}{|l|}{\textbf{LP}}&  \multicolumn{1}{|l|}{\textbf{TBCL}}&  \multicolumn{1}{|l|}{\textbf{TBCD}}&  \multicolumn{1}{|l|}{\textbf{TTCL}}&  \multicolumn{1}{|l|}{\textbf{TTCD}}&  \multicolumn{1}{|l|}{\textbf{Average}} 
    \\\hline
    \textbf{BiGRNN} & \textbf{MFCC} &  \textbf{0.8801} & \textbf{0.6200} & \textbf{0.8580} & \textbf{0.7382} & \textbf{0.6922} & \textbf{0.9206} & \textbf{0.7848} \\
    BiLSTM & MFCC &  0.8742 & 0.6236 & 0.8535 & 0.7189 & 0.6792 & 0.9142 & 0.7773 \\
    BiGRNN & MFCC +$\Delta+\Delta\Delta$ &  0.8708 & 0.6256 & 0.8565 & 0.7167 & 0.7020 & 0.9125 & 0.7807 \\
    BiLSTM & MFCC +$\Delta+\Delta\Delta$ &  0.8480 & 0.6112 & 0.8347 & 0.7055 & 0.6667 &  0.8975 & 0.7606 \\
    CNN-BiLSTM & Melspectrogram & 0.8285 & 0.5651 & 0.8028 & 0.6827 & 0.6193 & 0.8551& 0.7256
    \\\hline
  \end{tabular}}
\end{table}

\vspace*{-10pt}
\subsection{Effectiveness of different types of audio data augmentations}
\label{data_aug_results}
\vspace*{-2pt}

\begin{figure}
    \centering
\resizebox{220pt}{!}{%
    \begin{tikzpicture}[
      declare function={
        barW=4pt; 
        barShift=barW/2; 
      }
    ]
    \pgfplotsset{%
    height=0.38\textwidth,
    width=.7\textwidth
    }
    \begin{axis}[
        ybar = 0.6pt,
        ylabel=PPMC,
        ymin = 0.6,
        enlarge x limits = 0.14,
        bar width=9pt,
        legend style={at={(0.51,1.2)},
        anchor=north,legend columns=0},
        symbolic x coords={LA, LP, TBCL, TBCD, TTCL, TTCD, Average},
        xtick=data,
        legend image code/.code={
        \draw [#1] (0cm,-0.1cm) rectangle (0.2cm,0.25cm); }]
    
        \addplot[ybar,fill=blue] coordinates {
            (LA, 0.8872)
            (LP, 0.6456)
            (TBCL, 0.8619)
            (TBCD, 0.7504)
            (TTCL, 0.7091)
            (TTCD, 0.9213)
            (Average, 0.7959)
        };
        \addplot[ybar,fill=green] coordinates {
        (LA, 0.8810)
        (LP, 0.6321)
        (TBCL, 0.8567)
        (TBCD, 0.7288)
        (TTCL, 0.70523)
        (TTCD, 0.9210)
        (Average, 0.7875)
    };
    
        \addplot[ybar,fill=red] coordinates {
        (LA, 0.8809)
        (LP, 0.6464)
        (TBCL, 0.8631)
        (TBCD, 0.7408)
        (TTCL, 0.7035)
        (TTCD, 0.9194)
        (Average, 0.7924)
    };
    
    \legend{Background Noise,Gaussian Noise ,Room IR}
    \end{axis}
    \end{tikzpicture}
}
    \vspace*{-8pt}
    \caption{PPMC scores for each TV and the average score from the proposed BiGRNN model on clean speech test set}
    \label{barchart: augmentation_noises_ext}
\end{figure}
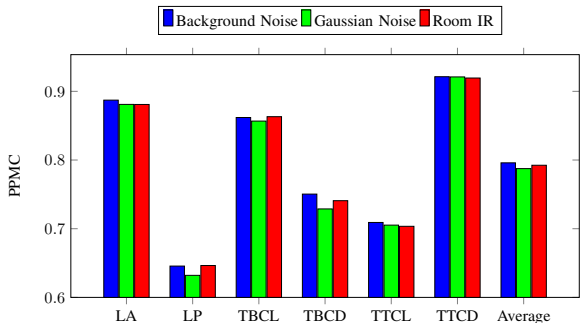

One of the key contributions of this work is the use of audio data augmentations to improve the standard SI task. But to choose the right data augmentation type, the best performing BiGRNN SI system was independently trained with speech data augmented with the three proposed data augmentation techniques in section \ref{data_augmentation}. The individual SI systems were then independently tested with the clean speech test split. Figure \ref{barchart: augmentation_noises_ext} shows the PPMC scores for each TV, and the resulting average score from the BiGRNN model trained and evaluated in this fashion. Results suggest that adding background noise and music from MUSAN dataset performs slightly better compared to adding reverberation noise from IR functions or random Gaussian noise.

A second experiment was also done to evaluate each SI system trained with one augmentation type, tested with the data augmented with the other two types. As expected, from Table \ref{table:augmentation_comp}, it is clear that the models trained with a certain augmentation type tends to perform well with the data augmented in the same fashion. Additionally, an average PPMC scores across different augmentation splits was also computed as shown in Table \ref{table:augmentation_comp}. The average PPMC scores further support the fact that adding background noise and music is better choice compared to the other two data augmentation types considered. 

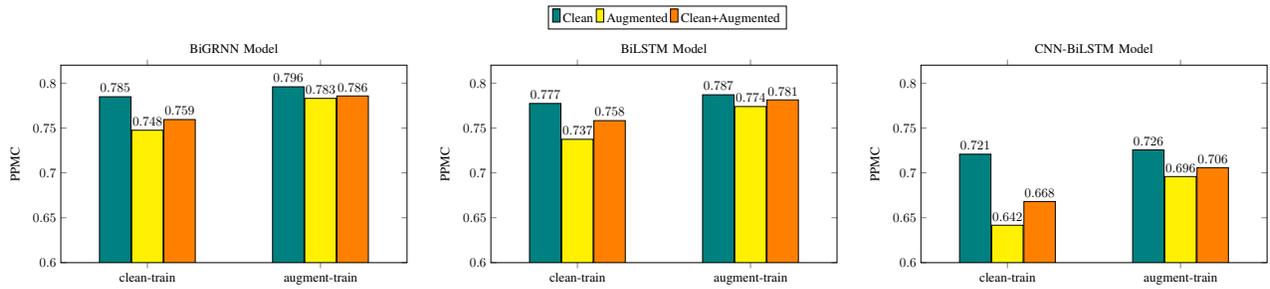
\begin{figure*}[h]
    \centering
\resizebox{488pt}{!}{%
\begin{tabular}{c c c}
\resizebox{200pt}{!}{%
    \begin{tikzpicture}[
        declare function={
        barW=4pt; 
        barShift=barW/2; 
      }
    ]
    \pgfplotsset{%
    height=0.38\textwidth,
    width=.6\textwidth
    }
    \begin{axis}[
        title = BiGRNN Model,
        ybar = 0.6pt,
        ylabel=PPMC,
        enlarge x limits = 0.5,
        ymin = 0.6, ymax = 0.82,
        bar width=24pt,
        nodes near coords={\pgfmathprintnumber[fixed zerofill, precision=3]{\pgfplotspointmeta}},
        legend style={at={(0.45,-0.15)},
        anchor=north,legend columns=0},
        symbolic x coords={clean-train, augment-train},
        xtick=data]
    
        \addplot[ybar,fill=teal] coordinates {
            (clean-train,0.7848)(augment-train,0.7959)
        };
        \addplot[ybar,fill=yellow] coordinates {
            (clean-train,0.7475)(augment-train,0.7830)
        };
        \addplot[ybar,fill=orange] coordinates {
        (clean-train,0.7594)(augment-train,0.7857)
        };
    
    \end{axis}
    \end{tikzpicture}
}
&
\resizebox{200pt}{!}{%
    \begin{tikzpicture}[
        declare function={
        barW=4pt; 
        barShift=barW/2; 
      }
    ]
    \pgfplotsset{%
    height=0.38\textwidth,
    width=.6\textwidth
}
    \begin{axis}[
        title = BiLSTM Model,
        ybar = 0.6pt,
        ylabel=PPMC,
        enlarge x limits = 0.5,
        ymin = 0.6, ymax = 0.82,
        bar width=24pt,
        nodes near coords={\pgfmathprintnumber[fixed zerofill, precision=3]{\pgfplotspointmeta}},
        legend style={at={(0.51,1.3)},
        anchor=north,legend columns=0},
        symbolic x coords={clean-train, augment-train},
        xtick=data,
        legend image code/.code={
        \draw [#1] (0cm,-0.1cm) rectangle (0.2cm,0.25cm); }]
    
        \addplot[ybar,fill=teal] coordinates {
            (clean-train,0.7773)(augment-train,0.7870)
        };
        \addplot[ybar,fill=yellow] coordinates {
            (clean-train,0.7374)(augment-train,0.7739)
        };
        \addplot[ybar,fill=orange] coordinates {
        (clean-train,0.7582)(augment-train,0.7813)
        };
        \legend{Clean, Augmented, Clean+Augmented}

    \end{axis}
    \end{tikzpicture}
}
&
\resizebox{200pt}{!}{%
    \begin{tikzpicture}[
        declare function={
        barW=4pt; 
        barShift=barW/2; 
      }
    ]
    \pgfplotsset{%
    height=0.38\textwidth,
    width=.6\textwidth
    }
    \begin{axis}[
        title = CNN-BiLSTM Model,
        ybar = 0.6pt,
        ylabel=PPMC,
        enlarge x limits = 0.5,
        ymin = 0.6, ymax = 0.82,
        bar width=24pt,
        nodes near coords={\pgfmathprintnumber[fixed zerofill, precision=3]{\pgfplotspointmeta}},
        legend style={at={(0.45,-0.15)},
        anchor=north,legend columns=0},
        symbolic x coords={clean-train, augment-train},
        xtick=data,
        legend image code/.code={
        \draw [#1] (0cm,-0.1cm) rectangle (0.2cm,0.25cm); }]
    
        \addplot[ybar,fill=teal] coordinates {
            (clean-train,0.7208)(augment-train,0.7256)
        };
        \addplot[ybar,fill=yellow] coordinates {
            (clean-train,0.6417)(augment-train,0.6959)
        };
        \addplot[ybar,fill=orange] coordinates {
        (clean-train,0.6681)(augment-train,0.7058)
        };
    
    \end{axis}
    \end{tikzpicture}
}
\end{tabular}}
\caption{Performance of each model with and without augmented data on clean, augmented and clean+augmented test sets}
\vspace*{-10pt}
\label{barchart: model_comparison}
\end{figure*}



\vspace*{-3pt}
\begin{table}[th]
  \caption{PPMC scores for models trained with one type of data augmentation, tested with the other types}
  \vspace*{-2pt}
  \centering
  \resizebox{\columnwidth}{!}{\begin{tabular}{|c| c| c| c| c|}
    \hline
    \multicolumn{1}{|c|}{\backslashbox{\textbf{Train Aug.}}{\textbf{Test Aug.}}} & 
                                    \multicolumn{1}{c|}{\textbf{\begin{tabular}{@{}c@{}}\textbf{Background} \\ \textbf{Noise \& Music} \end{tabular}}}  &      \textbf{Gaussian Noise}  &  \textbf{Room IR} & \textbf{Average} \\
    \hline
 
    Background Noise \& Music & \textbf{0.7830} & 0.7735 & 0.7572 & \textbf{0.7712}\\
    Gaussian Noise & 0.7726 & \textbf{0.7843} & 0.7395 & 0.7655  \\
    Room IR & 0.7624 & 0.7337 & \textbf{0.7797} & 0.7586
    \\\hline
  \end{tabular}}
  \label{table:augmentation_comp}
\end{table}
\vspace*{-8pt}

\subsection{Performance of proposed data augmentations}

Since data augmentation with background noise and music resulted in the best SI models, we trained all the subsequent models with data augmented in that fashion. Figure \ref{barchart: model_comparison} shows the average PPMC score across the 6 TVs for the proposed BiGRNN, BiLSTM and CNN-BiLSTM models when trained with clean speech only (clean-train), and clean speech + augmented data (augment-train). The two SI systems for each model are tested with clean only, clean+augmented and augmented only test sets to evaluate the robustness of the models not only for noisy/augmented speech, but also for clean speech data. The best performing SI model is reported with the BiGRNN model, trained with augmented data, and it predicts TVs with an average PPMC score of 0.7959 on clean speech data. For the context, the previous noise-robust model in \cite{seneviratne18_interspeech} was only able to achieve a best PPMC score of 0.741 on the clean test set. Therefore, the new SI system gains a 5\% relative improvement over the previous noise-robust SI system on estimating TVs for clean speech data. This elucidates the fact that, data augmentation can be used as a promising technique not only for making SI systems noise-robust, but also for improving the general SI task on clean speech data.

\vspace*{-5pt}
\subsection{Model adaptation for unseen-target speakers}
\vspace*{-2pt}

In previous work with SI systems, it has been shown that model adaptation with a generalized (pre-trained) model trained with a larger number of subjects can perform better than a speaker-dependent model trained only with target speaker's data \cite{illa18_interspeech}. Based on that observation, we performed model adaptation for speakers in the test set to see how much of an improvement the models can gain when an already pre-trained model is further trained with a portion of the subject's data. 

We take the best performing BiGRNN model pre-trained with augmented data and further train the model with each speaker's data in the test set to create individual speaker adapted models. The pre-trained model weights are only used for initialization and the new training was carried out with a smaller batch size (=4), reduced starting LR (=1e-4) and a quick decay with LR scheduler (every 2 epochs). The early stopping patience was increased to 30 to account for slight fluctuations in the validation loss due to smaller amount of training data. Here we used 80 \% of all the subject's data (average ~12 mins of speech) for training and used 10 \% each for validation and testing. Table \ref{table:model_adaptation_1} shows the PPMC results of BiGRNN models when tested with same test split of subject's data before and after model adaptation. Turns out for the 5 speakers in the test set, an average PPMC score of 0.8506 can be achieved when the speaker-adapted models are tested with held-out data from the same subject. The pre-trained model without any adaptation, tested on the same splits of the target subjects can only achieve a PPMC score of 0.7942.

\begin{figure}[t]
  \centering
  \includegraphics[width=1.0\columnwidth, height=55mm]{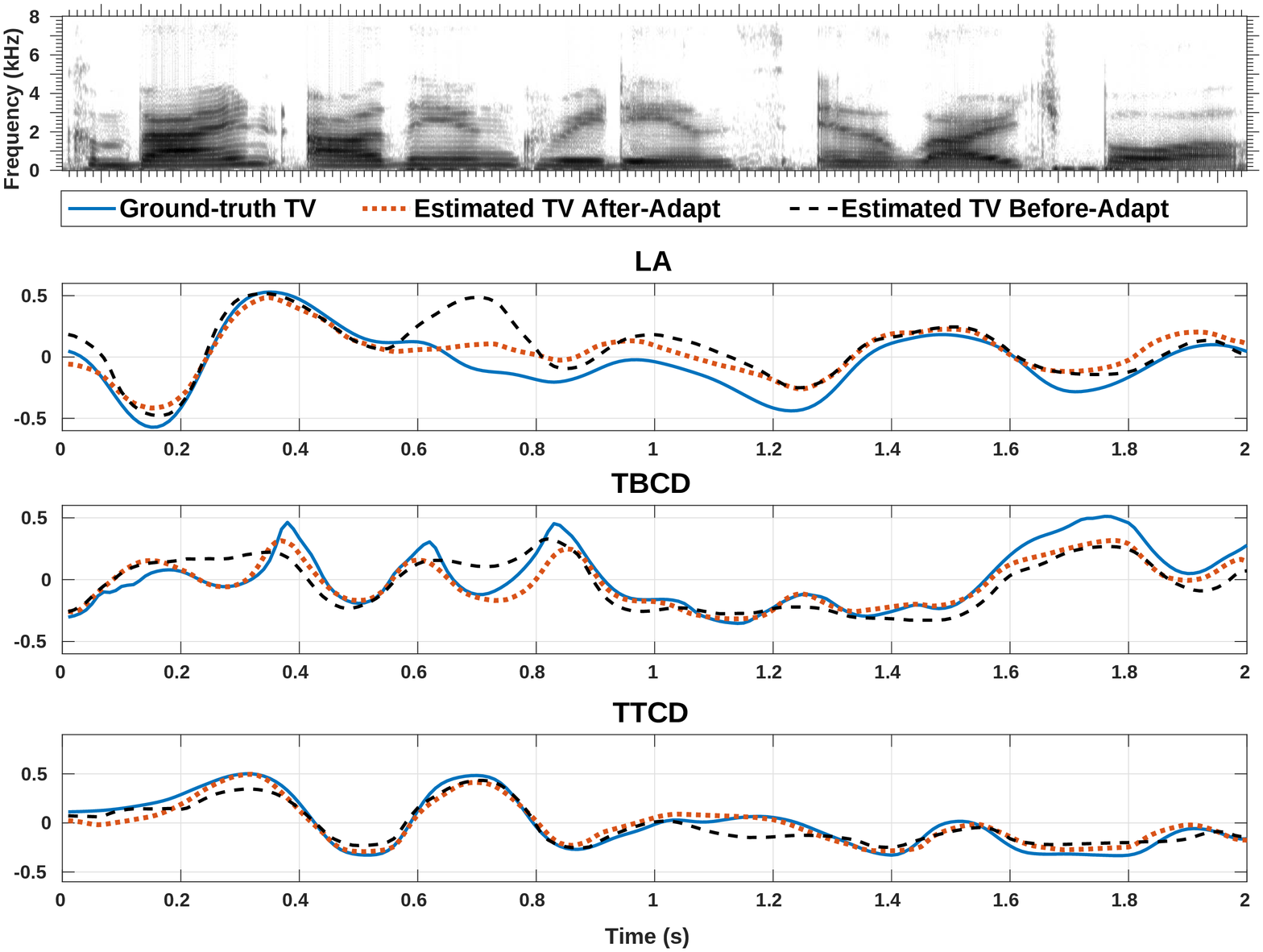}
  \caption{TV plots for the clean utterance `combined are the ingredients in a large bowl' estimated by the SI systems before and after adaptation on the JW31 subject}
  \vspace*{-12pt}
\label{fig:tv_plots}
\end{figure}

Figure \ref{fig:tv_plots} shows ground-truth and predicted TVs, LA,
TBCD, and TTCD for an example utterance of JW31 subject. The TVs are estimated by the pre-trained and the speaker-adapted SI systems. The effect of speaker adaptation is clearly evident with the estimated TV trajectories, where the predicted TVs from the speaker adapted model looks significantly better than that estimated by the generalized model. However, it should also be noted, that for certain speakers (e.g. JW61) this improvement is not clearly evident. Hence, further work needs to be done to understand what speaker-specific characteristics are captured by these speaker-adapted SI systems and to devise effective modifications to infuse speaker characteristics to improve speaker-adapted SI systems.

\vspace*{-8pt}
\begin{table}[th]
  \caption{PPMC score for TV predictions, before and after model adaptation}
  \vspace*{-8pt}
  \centering
  \resizebox{\columnwidth}{!}{\begin{tabular}{r r r}
    \\\toprule
    \multicolumn{1}{c}{\textbf{Subject}} & 
                                    \multicolumn{1}{c}{\textbf{Before Model Adaptation}}  &      \multicolumn{1}{c}{\textbf{After Model Adaptation}} \\
    \midrule
 
    \label{table: augmentation_comparison}
    \textbf{JW31} & 0.8071 & \textbf{0.9106}\\
    \textbf{JW39} & 0.7907 & \textbf{0.8836}\\
    \textbf{JW18} & 0.7936 & \textbf{0.8633}\\
    \textbf{JW33} & 0.8334 & \textbf{0.8490} \\
    \textbf{JW61} & 0.7463 & \textbf{0.7466}
    \\\bottomrule
  \end{tabular}}
  \vspace*{-5pt}
  \label{table:model_adaptation_1}
\end{table}

\vspace*{-5pt}

\vspace*{-2pt}
\section{Conclusion and Future Work}
\vspace*{-2pt}

In this work, we propose a BiGRNN model architecture to perform acoustic-to-articulatory speech inversion on the XRMB dataset. We show that audio augmentations can noticeably improve the SI task to perform well in both noisy and clean speech data. We also compare our proposed model with two baseline model architectures widely used in the SI domain. Finally, with speaker adaptation, we show that the performance of the BiGRNN model can be further improved for unseen speakers. As future work, we are hoping to conduct experiments with multiple articulatory corpora to explore how well data augmentation with multiple datasets can help in better generalizability of SI systems.

\vspace{12pt}

\end{document}